\documentclass[doublespacing]{elsart}

\usepackage{icarus}

\usepackage{natbib}

\bibpunct{(}{)}{;}{a}{,}{,}

\usepackage{graphicx}
\usepackage{lscape}
\usepackage{longtable}
\usepackage[modulo, pagewise]{lineno}

\newcommand{\BibTeX}{ \textrm{B\kern-.05em\textsc{i\kern-.025em b}\kern-.08em
    T\kern-.1667em\lower.7ex\hbox{E}\kern-.125emX} }

\newcounter{ionctr}
\ifx \undefined \ion \newcommand{\ion}[2]{\setcounter{ionctr}{#2}{#1$\;${\small\rmfamily\Roman{ionctr}}\relax}} \fi

\begin{document}

\begin{frontmatter}



\title{The CO$_2$ Abundance in Comets C/2012 K1 (PanSTARRS), C/2012 K5 (LINEAR), and 290P/J\"{a}ger as Measured with \textit{Spitzer}}


\author[label1]{Adam J. McKay},
\author[label2]{Michael S. P. Kelley},
\author[label1]{Anita L. Cochran},
\author[label2]{Dennis Bodewits},
\author[label3,label4]{Michael A. DiSanti},
\author[label5]{Neil Dello Russo},
\author[label5]{Carey M. Lisse}

\address[label1]{Univerisity of Texas Austin/McDonald Observatory, 1 University Station, Austin, TX 78712, (U.S.A);amckay@astro.as.utexas.edu, anita@astro.as.utexas.edu}
\address[label2]{Department of Astronomy, University of Maryland, College Park, MD 20742-2421 (U.S.A.); msk@astro.umd.edu, dennis@astro.umd.edu}
\address[label3]{NASA Goddard Center for Astrobiology, NASA GSFC, Mail Stop 690, Greenbelt, MD 20771 (U.S.A.); Michael.A.Disanti@nasa.gov}
\address[label4]{Solar System Exploration Division, Mail Stop 690, Greenbelt, MD 20771 (U.S.A)}
\address[label5]{Johns Hopkins University Applied Physics Laboratory, 11100 Johns Hopkins Rd., Laurel, MD, 20723 (U.S.A.); neil.dello.russo@jhuapl.edu, carey.lisse@jhuapl.edu}

\begin{center}
\scriptsize
Copyright \copyright\ 2015 Adam J. McKay, Michael S. Kelley, Anita L. Cochran, Dennis Bodewits, Michael A. DiSanti, Neil Dello Russo, Carey M. Lisse
\end{center}


%
%
%
%
%


\end{frontmatter}



\begin{flushleft}
\vspace{1cm}
Number of pages: \pageref{lastpage} \\
Number of tables: \ref{lasttable}\\
Number of figures: \ref{lastfig}\\
\end{flushleft}


\begin{pagetwo}{CO$_2$ in Comets PanSTARRS, LINEAR, and J\"{a}ger}

Adam J. McKay \\
University of Texas Austin\\
2512 Speedway, Stop C1402\\
Austin, TX 78712, USA. \\
\\
Email: amckay@astro.as.utexas.edu\\
Phone: (512) 471-6493

\end{pagetwo}

\begin{abstract}
Carbon dioxide is one of the most abundant ices present in comets and is therefore important for understanding cometary composition and activity.  We present analysis of observations of CO$_2$ and [\ion{O}{1}] emission in three comets to measure the CO$_2$ abundance and evaluate the possibility of employing observations of [\ion{O}{1}] emission in comets as a proxy for CO$_2$.  We obtained NIR imaging sensitive to CO$_2$ of comets C/2012 K1 (PanSTARRS), C/2012 K5 (LINEAR), and 290P/J\"{a}ger with the IRAC instrument on \textit{Spitzer}.  We acquired observations of [\ion{O}{1}] emission in these comets with the ARCES echelle spectrometer mounted on the 3.5-meter telescope at Apache Point Observatory and observations of OH with the \textit{Swift} observatory (PanSTARRS) and with Keck HIRES (J\"{a}ger).  The CO$_2$/H$_2$O ratios derived from the \textit{Spitzer} images are 12.6 $\pm$ 1.3\% (PanSTARRS), 28.9 $\pm$ 3.6\% (LINEAR), and 31.3 $\pm$ 4.2\% (J\"{a}ger).  These abundances are derived under the assumption that contamination from CO emission is negligible.  The CO$_2$ abundance for PanSTARRS is close to the average abundance measured in comets at similar heliocentric distance to date, while the abundances measured for LINEAR and J\"{a}ger are significantly larger than the average abundance.  From the coma morphology observed in PanSTARRS and the assumed gas expansion velocity, we derive a rotation period for the nucleus of about 9.2 hours.  Comparison of H$_2$O production rates derived from ARCES and \textit{Swift} data, as well as other observations, suggest the possibility of sublimation from icy grains in the inner coma.  We evaluate the possibility that the [\ion{O}{1}] emission can be employed as a proxy for CO$_2$ by comparing CO$_2$/H$_2$O ratios inferred from the [\ion{O}{1}] lines to those measured directly by \textit{Spitzer}.  We find that for PanSTARRS we can reproduce the observed CO$_2$ abundance to an accuracy of $\sim$ 20\%.  For LINEAR and J\"{a}ger, we were only able to obtain upper limits on the CO$_2$ abundance inferred from the [\ion{O}{1}] lines.  These upper limits are consistent with the CO$_2$ abundances measured by \textit{Spitzer}.
\end{abstract}

\begin{keyword}
Comets; Comets, Coma; Comets, Composition
\end{keyword}


\section{Introduction}
\indent The abundances of CO and CO$_2$ in comets may either reflect thermal evolution~\citep{BeltonMelosh2009} or formation conditions~\citep{AHearn2012}, with the distinct possibility that both evolution and formation conditions play a major role.  The formation of CO$_2$ likely occurs via grain surface interactions of OH and CO, though this reaction is not completely understood~\citep[e.g.][]{GarrodPauly2011,Noble2011}.  Another possible pathway is direct oxidation of CO on grain surfaces~\citep{Minissale2013}.  In either case, this would imply that CO$_2$ forms from destruction of CO and hence, if these reactions are efficient (i.e. most CO in the protosolar disk is converted to CO$_2$ via these reactions), on average CO$_2$ should be more abundant than CO in comets.  This abundance pattern could also be caused by thermal evolution due to CO being more volatile than CO$_2$ or the protosolar disk inherently having a CO$_2$/CO ratio greater than unity in the comet forming region that was inherited from the ISM.  In the case of evolution there should be observed trends in CO/CO$_2$ ratios as a function of the dynamical history of the comet. Definitive evidence for any such trend has not been observed~\citep{AHearn2012}, though it is possible that not enough comets have been observed for any trend that is present to become apparent.  Therefore, knowledge of the CO and CO$_2$ abundances in comets is paramount for creating a complete picture of cometary composition and differentiating between the effect of formation conditions and subsequent thermal evolution on cometary composition.\\

\indent However, the lack of a permanent dipole moment for CO$_2$ means it is best observed directly through its vibrational transitions at infrared wavelengths (CO$_2$ also has electronic transitions, but these are very weak and have never been observed astronomically).  The only successful direct observations of CO$_2$ in comets have been of its $\nu_3$ vibrational band at 4.26 $\mu$m, which is heavily obscured by the presence of telluric CO$_2$ and therefore cannot be observed from the ground.  Before 2004, the CO$_2$ abundance had been measured for only a few comets~\citep{Combes1988,Crovisier1997}, with many more observations becoming available over the last 10 years thanks to space-borne assets.  Observations of CO$_2$ in comets by \textit{Spitzer}~\citep{Pittichova2008,Reach2009, Reach2013}, AKARI~\citep{Ootsubo2010,Ootsubo2012}, \textit{WISE}~\citep{Bauer2011, Bauer2012, Stevenson2015,Bauer2015}, and the Deep Impact spacecraft~\citep{Feaga2007, AHearn2011, Feaga2014}, have revealed that CO$_2$ is the second most abundant gas present in most cometary comae (behind H$_2$O).  This may favor a mechanism where the CO$_2$ present in comets was formed via reactions that destroy CO and also possibly favor the idea that the measured abundances are indeed primordial.  Observations of Cameron band emission of CO with \textit{HST} have also been employed as a proxy for CO$_2$~\citep{Weaver1994}.  However, Cameron band emission has a significant contribution from electron impact excitation of CO, which complicates derivation of the CO$_2$ abundance~\citep{Bhardwaj2011}.  

\indent The only facilities that can currently observe CO$_2$ in comets are the \textit{Spitzer} Space Telescope and the \textit{WISE} spacecraft (directly in the IR), as well as \textit{HST} (through Cameron band emission.)  However, as these are all space-borne facilities, the observing time available is limited.  In addition, \textit{Spitzer} and \textit{WISE} all have very stringent elongation requirements, meaning many objects go unobserved (even though \textit{WISE/NEOWISE} is a survey, it only observes at 90$^{\circ}$ elongation, meaning observations of comets are serendipitous and cannot be planned for detailed study of a particular comet).  The James Webb Space Telescope (JWST) is expected to supercede the capabilities provided by \textit{Spitzer} and \textit{WISE} for cometary science~\citep{Kelley2015}, but JWST observing time for comets may be limited.  Ground-based observations are in general more accessible than space-borne assets, allowing for more detailed study of a larger number of objects.  Therefore, establishment of an indirect, ground-based measure of CO$_2$ abundances in comets is vital in order to provide the number of measurements needed for further interpretation of comet origin and evolution.\\

\indent As atomic oxygen is a photodissociation product of CO$_2$, observations of the forbidden oxygen lines at 5577, 6300, and 6364~\AA~can serve as a viable proxy.  These forbidden lines are fairly bright features in cometary spectra and can be readily observed in moderately bright comets (V=10) with medium aperture telescopes (2-3 meter class)~\citep[][and references therein]{Capria2005, Cochran2008, Decock2013, McKay2015}.  However, the photochemistry of \ion{O}{1} release from CO$_2$ photodissociation, as well as from its other primary parents H$_2$O and CO, is still poorly understood, and limits the usefulness of \ion{O}{1} as a reliable proxy~\citep{McKay2013, Decock2013}.  However, if independent, contemporaneous measurements of H$_2$O, CO$_2$, CO, and \ion{O}{1} are available, it is possible to employ comets as a ``laboratory'' to constrain the relevant photochemistry.

\indent We present analysis of \textit{Spitzer} Infrared Array Camera (IRAC) imaging of comets C/2012 K1 (PanSTARRS), C/2012 K5 (LINEAR), and 290P/J\"{a}ger (hereafter PanSTARRS, LINEAR, and J\"{a}ger, respectively), which we employ to measure the CO$_2$ production rate in each comet.  We also present high resolution optical spectroscopy of these comets in an effort to observe the [\ion{O}{1}] emission and therefore constrain the photochemistry responsible for the release of \ion{O}{1} into the coma.  In section 2 we describe our observations and reduction and analysis procedures.  Section 3 presents our CO$_2$ production rates and [\ion{O}{1}] line measurements, and a comparison of the CO$_2$ abundances inferred from the [\ion{O}{1}] emission and the abundances measured with \textit{Spitzer}.  In section 4 we discuss how the measured CO$_2$ production rates fit in with the growing sample of CO$_2$ observations in comets, as well as the implications of our results for the photochemistry of \ion{O}{1} release and the ability to use [\ion{O}{1}] observations as a proxy for CO$_2$.  Section 5 summarizes our conclusions.

\section{Observations and Data Analysis}
\subsection{Observations}
\indent We obtained NIR images at 3.6 and 4.5 $\mu$m for studying CO$_2$ using the IRAC instrument on \textit{Spitzer}~\citep{Werner2004, Fazio2004}, while we obtained optical spectra for studying atomic oxygen with the ARCES echelle spectrometer mounted on the Astrophysical Research Consortium 3.5-m telescope at Apache Point Observatory (APO) in Sunspot, New Mexico.  We also obtained optical spectra of J\"{a}ger with the HIRES instrument mounted on Keck I and imaging of PanSTARRS with the \textit{Swift} spacecraft to measure the OH production rate, which gives us a measure of the H$_2$O production rate.\\
  
\subsubsection{CO$_2$ -  \textit{Spitzer} IRAC}  
\indent As \textit{Spitzer} is well into its post-cryogenic mission, IRAC presently observes in two pass bands: one centered at 3.6 $\mu$m and the other at 4.5 $\mu$m.  Both filters have broad wavelength coverage, with bandwidths of 0.8 and 1.0 $\mu$m, respectively.  The 4.5 $\mu$m band is very useful for measuring the CO$_2$ abundances in comets, as this pass band includes the $\nu_3$ transition at 4.26 $\mu$m.  It also contains the $\nu$(1-0) band of CO at 4.7 $\mu$m, but in 15 out of 17 comets in the AKARI survey~\citep{Ootsubo2012}, the CO$_2$ feature was at least 10 times brighter than the CO feature, and so CO$_2$ is typically the dominant gas emission feature in the IRAC 4.5 $\mu$m band.  This is due to the fluorescence efficiency of CO$_2$ being approximately an order of magnitude larger than that for CO, while the CO abundance in comets is typically equal to or less than the CO$_2$ abundance.  While this is true for most comets, there are examples, such as C/2006 W3 (Christensen) and 29P/Schwassman-Wachmann 1, where CO emission contributes significantly (more than 20\%) to the 4.5 $\mu$m band flux~\citep{Ootsubo2012,Reach2013}.\\

\indent We supply details of our observations in Table~\ref{Spitzer}.  The IRAC array is a 256 x 256 pixel InSb array, covering a 5' x 5' region on the sky.  We performed observations of each comet field several days after each cometary observation in order to image the field without the comet in it.  These observations are termed ``shadow observations'' and provide a measurement of the background to be subtracted from the cometary images.  We observed each comet in high dynamic range mode.  This entailed obtaining exposures with both short and long exposure times in order to avoid saturation of the inner coma, while still keeping high signal-to-noise ratio (SNR) in the fainter outer coma (details of the exposure times used are given in Table~\ref{Spitzer}).  Observing in high dynamic range mode also helps protect against saturation due to bright field stars.  For these observations no pixels were saturated in the longest exposure times, therefore we performed analysis on the longest exposure time images for optimal SNR.\\

\indent For each comet, we combined all images of the same exposure time using the MOPEX software \citep{Makovoz2005}.  This process creates a mosaic in the rest frame of the comet from the individual images, averaging overlapping data together, but ignoring
cosmic rays and bad pixels.  Two mosaics are created: one for the comet data, the other for the shadow (background) data.  We subtracted
the shadow mosaic from the comet mosaic to remove the background.  This includes zodiacal light and celestial sources.\\

\indent After the mosaic images were created and the sky background was separated, the next step was to remove the dust contribution from the 4.5 $\mu$m band flux, isolating the gas emission.  We accomplished this via the following method.  First we split the 4.5 $\mu$m band image into wedges centered on the optocenter of the comet (for PanSTARRS, this consituted 20 wedges, while for J\"{a}ger and LINEAR the best results were derived assuming spherical symmetry, i.e. no splitting of the images into wedges was applied).  We then fit the 4.5 $\mu$m band image morphology in each wedge (or in the case of LINEAR and J\"{a}ger, the whole image) to a model consisting of a 1/$\rho$ profile (to approximate the gas) plus the 3.6 $\mu$m band image (indicative of the dust).  Both the 1/$\rho$ profile (i.e. gas model) and 3.6 $\mu$m band fluxes were allowed to be multiplied by a scale factor.  The gas scale factor was allowed to vary from wedge to wedge, but the scaling factor for the 3.6 $\mu$m band was forced to be the same for all wedges.  The key parameter we retrieved from this modelling is the 3.6 $\mu$m band scale factor.  Lastly, we multiplied the 3.6 $\mu$m band image by the retrieved scale factor and subtracted the scaled 3.6 $\mu$m band image from the 4.5 $\mu$m band image to obtain a dust-subtracted image.\\

\indent From the dust-subtracted image, we measured the flux for apertures ranging from 6-60 pixels (7-70'') in radius.  We converted the broadband photometry to CO$_2$ line fluxes following the IRAC data handbook~\citep{Laine2015}.  The line fluxes were then used to calculate average column densities inside each aperture employing fluorescence efficiencies from~\cite{CrovisierEncrenaz1983}.  Then the production rate $Q$ is given by
\begin{equation}
Q=<N>vd 
\end{equation}
where $<N>$ is the average column density in the photometric aperture, $v$ is the expansion velocity, and $d$ is the projected diameter of the photometric aperture.  We assume an expansion velocity of the coma following~\cite{Tseng2007}:
\begin{equation}
v=0.96R_h^{-0.44} 
\end{equation}
where $R_h$ denotes the heliocentric distance in AU of the comet and the derived velocity is in km s$^{-1}$.  This approach assumes a negligible effect of photodissociation on the spatial profile in the photometric aperture, but as our apertures are $<$ 10\% of the CO$_2$ scale length, this approximation is justified.  We calculated production rates for a variety of aperture sizes to quantify any trends in derived production rates with aperture size.  We find that small trends with aperture size are present, which may be due to residual dust, structure in the gas coma (i.e., the assumed 1/$\rho$ relation for the gas surface brightness profile is violated), and/or the difference between the 3.6 and 4.5 $\mu$m point-spread functions.  Residual dust may be present due to color variations in the coma.  We prefer to keep our approach simple, and refrain from using additional free-parameters and assumptions to model these deviations.  The uncertainties adopted for the CO$_2$ production rates are either derived from the 3.6 $\mu$m model scale factor uncertainties (LINEAR and J\"{a}ger), or from the standard deviation of the derived production rates for all apertures (PanSTARRS), whichever is larger.\\ 

To enhance structures in the coma morphology, we divided each image by a 1/$\rho$ profile, where $\rho$ is the projected distance from the optocenter.  A 1/$\rho$ profile is what is expected for a coma in steady state expansion.  Any deviations from this theoretical spatial profile are enhanced in the resulting image, allowing studies of the coma morphology to be performed.\\

\subsection{\ion{O}{1} -  ARCES}
\indent ARCES is a cross-dispersed echelle spectrometer, providing a spectral resolution of R $\equiv$ $\frac{\lambda}{\Delta\lambda}$ = 31,500 and a spectral range of 3500-10,000~\AA~with no interorder gaps.  This large, uninterrupted spectral range allows for simultaneous observations of all three oxygen lines.  More specifics for this instrument are discussed elsewhere~\citep{Wang2003}.\\

\indent The observation dates and geometries are described in Table~\ref{observations}.  All nights were photometric, meaning absolute flux calibration of the spectra was possible.  We used an ephemeris generated from JPL Horizons for non-sidereal tracking.  Guiding was accomplished using a boresight technique, which utilizes optocenter flux that falls outside the slit to keep the slit on the optocenter.  We observed a G2V star, a fast rotating (vsin(i) $>$ 150 km s$^{-1}$) O, B, or A star, and a flux standard for calibration of the comet spectra.  The calibration stars used for each observation date are given in Table~\ref{observations}.  We obtained spectra of a quartz lamp for flat fielding and acquired spectra of a ThAr lamp for wavelength calibration.\\

\indent Spectra were extracted and calibrated using Image Reduction and Analysis Facility (IRAF) scripts that perform bias subtraction, cosmic ray removal, flat fielding, and wavelength calibration.  We employed the fast-rotator spectrum to remove telluric features, the flux standard spectrum to convert from counts to physical units, and the solar analog spectrum to remove Fraunhofer lines.  We assumed an exponential extinction law and extinction coefficients for APO when flux calibrating the cometary spectra.  More details of our reduction procedures can be found in~\cite{McKay2015} and references therein.\\

\indent Because of the small size of the ARCES slit, it is necessary to obtain an estimate of the slit losses to achieve an accurate flux calibration.  We find the transmittance through the slit by performing aperture photometry on the slit viewer images as described in~\cite{McKay2014}.  This introduces a 10\% uncertainty in our absolute flux calibration.\\

\indent The \ion{O}{1} lines are also present as a telluric emission feature, meaning a combination of high spectral resolution and large geocentric velocity (and therefore large Doppler shift) is needed to separate the cometary line from the telluric feature.  For all observations the telluric and cometary lines are well separated.  We fit the line profiles using the  Gaussian-fitting method described in~\cite{McKay2012}.  Emission from the C$_2$ $\Delta$v=-1 Swan band can also contaminate the cometary 5577~\AA~feature~\citep[e.g.][]{Decock2013}.  However, there is no trace of C$_2$ emission in any of the observed comets in the wavelength region surrounding the 5577~\AA~feature.  Therefore we consider any contamination from C$_2$ negligible.  The 6300~\AA~and 6364~\AA~lines are both transitions from the $^1$D to the $^3$P ground state, therefore the flux ratio reflects the branching ratio for these transitions of 3.0.  This means that the flux ratio of these lines is independent of the coma physics, and the expected value of 3.0 is well established by both theory and observation~\citep{SharpeeSlanger2006, CochranCochran2001, Cochran2008, McKay2012, McKay2013, Decock2013}.  As a check of our analysis procedures, we confirmed that we reproduced this ratio before proceeding with further analysis.\\

\indent With the measured line fluxes, we calculate the oxygen line ratio, defined as
\begin{equation}
R=\frac{I_{5577}}{I_{6300}+I_{6364}}
\end{equation}
where $I_y$ denotes the flux of line $y$.  The CO$_2$/H$_2$O ratio can be inferred from the oxygen line ratio using the following relation~\citep{McKay2012,Decock2013}:   
\begin{equation}
\frac{N_{CO_2}}{N_{H_2O}} = \frac{RW^{red}_{H_2O}-W^{green}_{H_2O}-W^{green}_{CO}\frac{N_{CO}}{N_{H_2O}}+RW^{red}_{CO}\frac{N_{CO}}{N_{H_2O}}}{W^{green}_{CO_2}-RW^{red}_{CO_2}}
\end{equation}
where $N$ is column density and $R$ is the oxygen line ratio.  The release rate $W$ is defined as 
\begin{equation}
W \equiv \tau^{-1}\alpha\beta
\end{equation}
where $\tau$ represents the photodissociative lifetime of the parent molecule, $\alpha$ is the yield into the excited state of interest, and $\beta$ represents the branching ratio for a given line out of a certain excited state.  This relation is derived by noting that the line flux contributed from each species is given by the product of column density $N$ and release rate $W$, substituting this into Eq. 3 and solving for $\frac{N_{CO_2}}{N_{H_2O}}$ (see~\cite{McKay2012} for more details).  We ignore the contribution of more complex oxygen-bearing molecules like H$_2$CO and CH$_3$OH as these species are less abundant than H$_2$O, CO$_2$, and CO and release oxygen through a multi-step process, making them very inefficient at contributing to the \ion{O}{1} population.  If the contribution of CO photodissociation to the \ion{O}{1} population is also considered negligible~\citep{RaghuramBhardwaj2014}, Eq. 4 simplifies to~\citep{McKay2013}:
\begin{equation}
\frac{N_{CO_2}}{N_{H_2O}} = \frac{RW^{red}_{H_2O}-W^{green}_{H_2O}}{W^{green}_{CO_2}-RW^{red}_{CO_2}}
\end{equation}
The results of Eq. 4 and 6 are independent of heliocentric distance.  For small fields of view, the column density ratio reflects the production rate ratio~\citep[see][and references therein for more details]{McKay2015}.

\indent We performed additional analysis accounting for preferential collisional quenching of $^1$D atoms (responsible for the 6300~\AA~and 6364~\AA~lines) as compared to $^1$S atoms (responsible for the 5577~\AA~line), which can be important for small fields of view or high production rates~\citep{BhardwajRaghuram2012, RaghuramBhardwaj2014, Decock2015}.  The oxygen line ratio employed in Eqs. 4 and 6 assumes the ratio was calculated using 6300~\AA~and 6364~\AA~line intensities that are unaffected by collisional quenching.  Since this may not be the case, the observed 6300~\AA~and 6364~\AA~line intensities need to be increased to account for the $^1$D atoms that were de-excited through collisions and so do not contribute to the 6300~\AA~and 6364~\AA~line intensities.  In order to account for this, we need to model the number density of the dominant collisional partner, H$_2$O.  Therefore an estimate of the H$_2$O production rate is needed.\\

\indent We determined H$_2$O production rates from our [\ion{O}{1}]6300~\AA~line observations by employing algorithms based on those used in~\cite{Morgenthaler2007} and~\cite{McKay2012}, which involves a Haser model modified to emulate the more physical vectorial model.  With an H$_2$O production rate in hand, we estimate the percentage of atoms lost to collisional quenching by employing the algorithms mentioned above to estimate the expected [\ion{O}{1}]6300~\AA~flux without collisional quenching.  The correction factor is then simply the expected flux without quenching divided by the observed flux.  More details concerning this method are presented in~\cite{McKay2015}.\\

\subsection{OH-\textit{Swift} and HIRES}
\indent The \textit{Swift} telescope~\citep{Gehrels2004} observed PanSTARRS on May 6 and 7, 2014 at R=2.04 AU from the Sun.  We employed the UVOT instrument~\citep{Mason2004, Roming2005} to obtain photometry of the comet.  UVOT's broadband filters provide a measure of the comet's water and dust production rates~\citep{Bodewits2014}. We obtained photometry using broadband V (central $\lambda$ 5468~\AA, FWHM 750~\AA) and UVW1 (central $\lambda$ 2600~\AA, FWHM 700~\AA) filters. We used the UVW1 filter to detect OH emission and the V-band filter to remove the contribution of continuum in the UVW1 filter. We scaled the V-band flux to the UVW1 filter by assuming the reflected dust continuum is a solar spectrum with no reddening.  We note that there is likely some contamination from C$_2$ Swan band emission in the V-band filter.  Correcting for the filter transmission at the relevant wavelengths, the measured fluxes can be converted into column densities using heliocentric distance and velocity dependent fluorescent efficiencies~\citep{SchleicherAHearn1988}. To derive water production rates, we compare the measured OH content of the coma with an OH distribution calculated using the vectorial model \citep{Festou1981, Combi2004}.  Most of the uncertainty in the derived production rates is introduced from the modeling, with a negligible contribution coming from photon noise.  We measured fluxes in several aperture sizes, and adopt the standard deviation of the derived production rates as our 1-sigma uncertainty.

\indent For J\"{a}ger, we obtained observations of OH with the HIRES instrument~\citep{Vogt1994} on Keck I in January 2014.  The HIRESb configuration provides observations of the OH $\Delta v$=0 band at 3080~\AA~.  We utilized the 0.86 $\times$ 7.0'' slit.  Observing procedures, reduction, and analysis of the data are very similar to ARCES.  For these observations the Full Moon was only six degrees away from J\"{a}ger, meaning a large amount of scattered moonlight was present in the spectra.  The additional strong continuum introduced from the scattered moonlight dominates the Poisson noise in the spectra and was difficult to remove completely.  We extract the band flux using a spectral fitting model very similar to that presented in~\cite{McKay2014}.  This flux is then converted to an H$_2$O production rate using a Haser model that has been modified to emulate the vectorial model (see~\cite{McKay2014} for more details) and the fluorescence efficiency from~\cite{SchleicherAHearn1988}.  The scale lengths for the Haser model are adopted from~\cite{CochranSchleicher1993}.  

\section{Results}
\indent We provide measured fluxes for CO$_2$ from our \textit{Spitzer} observations (including 3.6 $\mu$m image scale factors, see section 2.1.1), [\ion{O}{1}]6300~\AA~emission from our ARCES observations, and OH from our \textit{Swift} and Keck HIRES observations in Table~\ref{fluxes}.  All uncertainties are 1-sigma.

\subsection{H$_2$O Production Rates and Collisional Quenching Factors}
\indent We show a spectrum of J\"{a}ger showing the OH lines (from which we derived the H$_2$O production rate) in Fig.~\ref{OH}.  As discussed in Section 2.3, the noisy background is largely due to scattered sunlight from the Full Moon.  We present our H$_2$O production rates and collisional quenching correction factors in Table~\ref{QCO2}.  A small collisional quenching factor is required for the PanSTARRS data, while for LINEAR and J\"{a}ger the effect is negligible due to their much smaller H$_2$O production rates.  As we are using [\ion{O}{1}]6300 emission to derive H$_2$O production rates, it is desirable to have independent production rates determined via other methods to confirm that there is no systematic error being introduced by employing [\ion{O}{1}] emission.  

\indent PanSTARRS was by far the brightest of these comets, and therefore several other measurements of the H$_2$O production rate are available.  Gibb et al. (private communication) measured the H$_2$O production rate with NIRSPEC, and their value is consistent with our value of (4.35 $\pm$ 0.44) $\times$ 10$^{28}$ molecules s$^{-1}$ derived from [\ion{O}{1}]6300 emission to within $\sim$ 10\%.  From the \textit{Swift}/UVOT observations we derived a water production rate of (9.5 $\pm$ 0.8) $\times$ 10$^{28}$ molecules s$^{-1}$ in apertures between 50-200 arcsec (5.3 $\times$ 10$^4$ - 2.1 $\times$ 10$^5$ km at the comet).  Analysis of OH observations by~\cite{KnightSchleicher2014} and of Lyman-$\alpha$ emission by~\cite{Combi2014} derive similar production rates.  One possibility for the discrepency is that the H$_2$O production rate depends on rotational phase of the nucleus.  However, as our [\ion{O}{1}]6300 observations and the NIRSPEC observations occured on completely different nights, it is unlikely that both would have sampled the same part of the rotational variation.  In addition, this would imply that the observations of Schleicher,~\cite{Combi2014}, and our \textit{Swift} observations (which also occured on different dates) would have all sampled the same part of the rotational variation that was also distinct from that sampled by our [\ion{O}{1}]6300 observations and the NIRSPEC observations.  A more likely possibility is related to the fields of view (FOV) of the different telescope/instrument combinations employed.  Our [\ion{O}{1}]6300 observations and the NIRSPEC observations of Gibb et al. both employed narrow slits (projected FOV at the comet on the order of several thousand km), while the \textit{Swift},~\cite{Combi2014}, and Schleicher observations all used much larger FOV, on the order of tens to hundreds of thousands of kilometers.  A similar dependence of derived H$_2$O production rates with FOV was observed for C/2009 P1 (Garradd)~\citep{Combi2013,Bodewits2014,DiSanti2014,Feaga2014,McKay2015}.  This was interpreted as an extended source of icy grains that sublimated outside the FOV of slit-based spectroscopic measurements, but within the FOV of narrow band imaging observations.  A similar phenomenon could be applicable to PanSTARRS.  This raises the question of which H$_2$O production rate is appropriate to adopt for comparison to our [\ion{O}{1}] observations.  As our \ion{O}{1} observations have a small FOV, we will employ the H$_2$O production rate derived from small FOV observations for the analysis throughout the rest of this paper (but see Section 4.1 for more discussion on how this affects comparison to other observations).\\     

\indent There are no other sources of H$_2$O production rates available for J\"{a}ger or LINEAR.  We have observations of OH for J\"{a}ger, but these are two months after the [\ion{O}{1}] observations.  However, as the collisional quenching was determined to be negligible for LINEAR and J\"{a}ger, any systematic uncertainties in our H$_2$O production rate due to employing [\ion{O}{1}]6300 emission to obtain an H$_2$O production rate will have a negligible effect on our CO$_2$/H$_2$O ratios inferred from the oxygen line ratio.\\

\indent It is possible that systematic uncertainties in the H$_2$O production rate will also affect the \textit{Spitzer}-derived CO$_2$/H$_2$O ratios.  However, as discussed above, independent direct observations of H$_2$O using NIRSPEC are consistent with our adopted H$_2$O production rate for PanSTARRS.  For J\"{a}ger, our preferred H$_2$O production rate for comparison to the \textit{Spitzer} measurement of CO$_2$ is from HIRES using OH due to this observation being more contemporaneous with the \textit{Spitzer} observations than the [\ion{O}{1}] observations.  LINEAR has no independent measure of the H$_2$O production rate, but the agreement of H$_2$O production rates derived for PanSTARRS and J\"{a}ger to other methods gives us confidence that our derived H$_2$O production rate for LINEAR is accurate to the quoted uncertainties.

\subsection{CO$_2$ Production Rates and Coma Morphology}
\indent In Fig.~\ref{Spitzer_images} we show the \textit{Spitzer} IRAC images of PanSTARRS, LINEAR, and J\"{a}ger.  PanSTARRS was by far the brightest of the three comets observed, as is evident in the quality of the images.  For PanSTARRS even in the raw mosaics (i.e. no image enhancement or dust subtraction), it is evident that in the 4.5 $\mu$m image there is a diffuse, extended emission that is not present in the 3.6 $\mu$m image and is likely due to CO$_2$ or CO gas.  We present the dust-subtracted images in Fig.~\ref{Spitzer_gas}.\\  

\indent We present the derived CO$_2$ production rates and CO$_2$/H$_2$O ratios under the assumption of negligible CO emission in Table~\ref{QCO2}.  The quoted uncertainties include only stochastic noise and uncertainties with the modeling used to isolate the gaseous emission, and do not include any systematic error associated with any possible CO emission.  While in principal there is both emission from CO$_2$ and CO in the 4.5 $\mu$m image, NIRSPEC observations by Gibbs et al. (private communication) constrain the CO/H$_2$O ratio for PanSTARRS at $\sim$ 3\%.  At this abundance the contribution of CO emission to the 4.5 $\mu$m flux is minimal (on the order of 3\%) and we can assume, within our uncertainties, that all the gas emission we observe is due to CO$_2$ in this case.  There are no independent measurements of CO available in J\"{a}ger or LINEAR, meaning in these cases our CO$_2$ production rates could in fact be upper limits.  However, for most comets in the AKARI survey~\citep{Ootsubo2012}, the CO emission was much weaker than that from CO$_2$.  Therefore in general it is likely that our CO$_2$ production rates for J\"{a}ger and LINEAR are not contaminated by CO emission, but without direct, independent observations of CO we cannot be certain.  However, even with a CO/H$_2$O ratio as high as~$\sim$ 30\% (higher than any comet in the AKARI survey except C/2006 W3 (Christensen) and 29P/Schwassman-Wachmann 1 and higher than all comets observed from ground-based IR spectroscopy~\citep{MummaCharnley2011} except C/2009 P1 (Garradd)~\citep{Feaga2014,McKay2015}), the derived CO$_2$/H$_2$O abundances from the \textit{Spitzer} observations of LINEAR and J\"{a}ger only drop to about 23\% and 27\%, respectively.  Therefore, for our derived CO$_2$ abundances to change significantly, LINEAR and J\"{a}ger would have to have extremely abnormal CO/H$_2$O ratios ($>$ 100\%, only observed for comets C/2006 W3 (Christensen), 29P/Schwassman-Wachmann 1, and C/1995 O1 (Hale-Bopp)~\citep{Biver2002} at larger heliocentric distances than the comets studied here) as compared to the observed sample of comets~\citep{Ootsubo2012, AHearn2012}.  For LINEAR, the presence of gas emission in the 4.5 $\mu$m images was not obvious and the detection of CO$_2$ is sensitive to model assumptions employed to isolate the gas emission.  Therefore in this case our derived CO$_2$ production rate may be better interpreted as an upper limit.\\

\indent In Fig.~\ref{PanSTARRS_morph}, the top row shows (from left to right) the 3.6 $\mu$m, 4.5 $\mu$m, and dust-subtracted images of PanSTARRS.  The bottom row is the same, except these images have been divided by a 1/$\rho$ profile to show coma features.  Figs.~\ref{LINEAR_morph} and~\ref{Jager_morph} show the analogous figures for LINEAR and J\"{a}ger, respectively.  All the images show a clear tail excess.  Even the gas images show some residual tails, suggesting that the dust subtraction is not perfect.  For PanSTARRS, a spiral shape is visible in the 4.5 $\mu$m image and the gas image, but is not present in the 3.6 $\mu$m image.  This is likely the manifestation of a CO$_2$ jet.  Observations of the CN morphology also show this spiral structure, while observations of the dust through R-band imaging do not~\citep{KnightSchleicher2014}.  The \textit{Swift} imaging of OH also does not show any discernible morphology.  This may indicate that the OH (and its parent H$_2$O) is released from the nucleus in a manner that is different from the CN parent and CO$_2$, but this could also be due to any morphology that is present being blurred by the random direction of the velocity that OH receives after photodissociation of H$_2$O.\\

\indent We can use the separation between arcs of the spiral morphology to obtain an estimate of the rotation period.  To determine the positions of the arcs, we measured the total flux in concentric annuli centered on the comet photocenter.  Annuli containing the arcs will have higher flux than adjacent annuli.  To increase the accuracy of the derived arc positions, we then fit the spatial distribution of flux within the annuli containing an arc with a Gaussian function plus constant background.  We derive mean peak centers at 25.1 $\pm$ 0.3 and 48.3$\pm$ 1.0 pixels from the optocenter, corresponding to 26700 $\pm$ 300 and 51200 $\pm$ 1100 km.  These positions predict additional arcs should be present at 1.9 and 71.5 pixels from the optocenter.  The 1.9 pixel offset is too close to the comet to resolve, but the surface brightness profile does seem to be peaked quite close to the center.  The 71.5 pixel peak is not apparent, but may be too diffuse to resolve in our data.  Assuming an expansion speed of 0.74 km/s in the plane of the sky (derived from the~\cite{Tseng2007} relation, see Section 2.1), the apparent period is 9.2 $\pm$ 0.4 hr, which we identify as a candidate rotation period for the nucleus.  The CN morphology observed by~\cite{KnightSchleicher2014} is consistent with this rotation period, but their analysis does not provide a definitive value.\\

\indent The spatial profile for all the LINEAR images is very symmetric and shows no notable features.  The J\"{a}ger dust-subtracted image shows possible asymmetry to the bottom of the image, but no obvious coma structures such as observed for PanSTARRS that could be used to derive a rotation period are present.\\

\subsection{\ion{O}{1} Line Ratios and Inferred CO$_2$ Abundances}
\indent We present our oxygen line ratio measurements and 3-sigma upper limits in Table~\ref{comparerates}.  Unfortunately, LINEAR and J\"{a}ger were not bright enough for detection of the [\ion{O}{1}]5577~\AA~line, and the upper limits are not particularly constraining.  However, PanSTARRS was much brighter and we have a firm detection of the [\ion{O}{1}]5577~\AA~line, as shown in Fig.~\ref{OIline}.\\

\indent We derive CO$_2$/H$_2$O ratios from our oxygen line ratios (or 3-sigma upper limits in the case of LINEAR and J\"{a}ger) using release rates from~\cite{BhardwajRaghuram2012} and~\cite{McKay2015}.  We summarize our CO$_2$ abundances directly measured by \textit{Spitzer} and our inferred CO$_2$ abundances from our oxygen line observations in Table~\ref{comparerates}.  The specific values for the release rates are given in Table~\ref{Wrates}.  The rates from~\cite{BhardwajRaghuram2012} are derived from a photochemical model of a cometary coma, while the empirical rates from~\cite{McKay2015} are rates that are able to reproduce the CO$_2$/H$_2$O ratio determined by~\cite{Feaga2014} for comet C/2009 P1 (Garradd).  The difference between empirical release rates A and B from~\cite{McKay2015} is a factor of 1.5 in the CO$_2$ release rates that accounts for differences in the CO/H$_2$O abundance in Garradd measured by~\cite{McKay2015} and~\cite{Feaga2014}.  For PanSTARRS, we used Eq. 4, which includes the contribution of CO, with a CO/H$_2$O ratio of $\sim$ 3\% (Gibbs et al. private communication).  As no independent measure of the CO abundance is available for J\"{a}ger or LINEAR, we applied Eq. 6, which assumes no contribution to the \ion{O}{1} population from CO.  If the contribution of CO is significant, this would not affect our upper limit, since including a contribution from CO only lowers the inferred upper limit on CO$_2$.  Therefore our derived upper limits are true upper limits.\\

Using release rates from~\cite{BhardwajRaghuram2012}, we infer a CO$_2$/H$_2$O ratio of $\sim$4\% for PanSTARRS, while the abundance measured by \textit{Spitzer} is approximately 12\%.  The empirical release rates from~\cite{McKay2015} reproduce the CO$_2$/H$_2$O ratio to better accuracy, predicting a CO$_2$/H$_2$O ratio of $\sim$10\% (release rates A) or $\sim$14\% (release rates B).  The upper limits inferred for J\"{a}ger and LINEAR using the~\cite{McKay2015} release rates are consistent with the values measured by \textit{Spitzer}, but do not provide further constraints on their accuracy.  The upper limit inferred for LINEAR using release rates from~\cite{BhardwajRaghuram2012} may be inconsistent with the \textit{Spitzer} result (see section 4.2), but the J\"{a}ger results are consistent with the~\cite{BhardwajRaghuram2012} release rates.\\

\section{Discussion}
\subsection{CO$_2$ Abundances}
\indent The CO$_2$/H$_2$O ratio of 12\% measured for PanSTARRS is slightly lower than the mean of the AKARI survey of comets measured at heliocentric distances of less than 2.5 AU (i.e. inside the canonical water sublimation line where sublimation rates of H$_2$O, CO$_2$, and CO do not vary much with respect to each other~\cite[e.g.][]{MeechSvoren2004}), which is approximately 17\%~\citep{Ootsubo2012}, but is well within the spread of values observed in comets at similar heliocentric distance to date.  The observed CO$_2$ abundances of 29\% and 31\% for LINEAR and J\"{a}ger, respectively, are higher than any comet observed by AKARI within 2.5 AU from the Sun.  However, these values are close to the mean value of 30\% found by~\cite{Reach2013}, although this data set has much more scatter than the AKARI survey.\\ 

\indent As both LINEAR (1.6 AU) and J\"{a}ger (2.2 AU) were observed at heliocentric distances less than 3 AU, sublimation effects are not likely responsible for these high abundances~\citep{MeechSvoren2004}. One possibility is that since AKARI observed over a much larger FOV (1'$\times$1') than our narrow slit observations, the AKARI observations were sensitive to any extended sources of H$_2$O that may have been present around the comets in their survey.  If an extended source of water was present in J\"{a}ger and LINEAR, this would have resulted in larger derived H$_2$O production rates and therefore smaller CO$_2$/H$_2$O ratios than would have been measured using H$_2$O prodution rates from narrow slit observations such as ours.  Therefore if we had used an observational set up similar to AKARI with a large FOV to derive H$_2$O production rates, our derived CO$_2$/H$_2$O ratios might have been be lower, bringing the measured abundances for LINEAR and J\"{a}ger closer to the mean value derived from AKARI.  A similar effect is expected for the~\cite{Reach2013} sample, as they adopted H$_2$O production rates from wide field OH imaging.  However, without any data indicating the magnitude of an extended source of H$_2$O production around LINEAR and J\"{a}ger, we cannot evaluate this possibility further.  Another caveat to consider is the possibility that CO emission is contributing to \textit{Spitzer}'s 4.5 $\mu$m filter, meaning the CO$_2$ production rates are in fact lower than presented here.  However, as discussed in Section 3.2, this is not likely, as an abnormally large CO abundance is required to change the derived CO$_2$/H$_2$O ratio significantly.  As mentioned in section 3.2, we cannot rule out the possibility that our detection of CO$_2$ in LINEAR is better interpreted as an upper limit, in which case its CO$_2$ abundance would be more typical.\\

\indent This study, while only adding three new comets to the sample, is consistent with the findings of previous CO$_2$ surveys in comets that the average CO$_2$ abundance in comets is about 15-30\%, higher than previously thought~\citep{Ootsubo2012,Reach2013}.  Only one of our comets has a measurement of the CO abundance (PanSTARRS), and the preliminary CO abundance in this comet derived by Gibbs et al. (private communication) is much less than the CO$_2$ abundance, consistent with the idea that the formation of CO$_2$ via grain-surface reactions involving CO is a viable pathway for CO$_2$ formation in the protosolar disk.  With no observations of CO in LINEAR or J\"{a}ger available, we cannot reach any conclusions on the CO/CO$_2$ ratio in those comets.\\

\subsection{Accuracy of \ion{O}{1} as a Proxy for CO$_2$}

\indent Using our \textit{Spitzer} observations, we were able to compare actual CO$_2$ abundances for these comets to CO$_2$ abundances inferred using observations of the oxygen line ratio.  For PanSTARRS, the~\cite{BhardwajRaghuram2012} release rates underestimate the CO$_2$ abundance by about a factor of three.  A similar discrepency was found for C/2009 P1 (Garradd)~\citep{McKay2015}.  The upper limits inferred from the oxygen line ratio for J\"{a}ger are not particularly constraining, as all three sets of release rates provide upper limits consistent with the \textit{Spitzer} measurements.  However, for LINEAR the upper limit on CO$_2$/H$_2$O using the~\cite{BhardwajRaghuram2012} release rates is similar to the value measured by \textit{Spitzer}.  While not conclusive, this suggests that we should have been able to detect the [\ion{O}{1}]5577~\AA~line in this comet, which we did not.  However, a lack of knowledge of the CO abundance and the quality of the data prevent us from making a firm conclusion.  Therefore there is suggestive (but not conclusive) evidence that the~\cite{BhardwajRaghuram2012} release rates do not reproduce the LINEAR observations.

\indent The empirical release rates from~\cite{McKay2015} also reproduce the CO$_2$ abundance observed in comet Garradd (by definition, as this was a requirement in the derivation of these release rates; the two sets of release rates correspond to different values of the CO/H$_2$O ratio used in Eq. 4).  The ability of these release rates to reproduce the CO$_2$ abundance in PanSTARRS to within an accuracy of 20\% is encouraging, but the lack of detections of the [\ion{O}{1}]5577 line in J\"{a}ger and LINEAR prevent further evaluation.  The empirical release rates B seem to reproduce the CO$_2$ abundance in PanSTARRS more accurately than the release rates A.  More simultaneous observations of CO$_2$ and the oxygen line ratios in comets are needed to further evaluate this method, and specifically the release rates proposed by~\cite{McKay2015}.\\

\indent It is important to stress that the release rates from~\cite{McKay2015} are strictly empirical.  They seem to satisfactorily reproduce current observations, but there is no physical explanation for why they are different from those derived using photochemical models, such as those presented in~\cite{BhardwajRaghuram2012}.  Laboratory measurements of the \ion{O}{1} release rates are required to help settle this discrepancy.  It may be possible that the release rates from~\cite{McKay2015} are simply effective release rates.  The release rates derived using photochemical models might be correct in the strict sense, but perhaps other physical processes occur in the coma (collisional processes, radiative transfer effects, etc.) that modify the [\ion{O}{1}] emission so that applying those release rates to remote sensing observations does not reproduce the measured CO$_2$ abundance of the comet.  A more detailed understanding of the coma environment and its effect on [\ion{O}{1}] emission is needed.\\

\section{Conclusions}
\indent We have presented near-contemporaneous observations of CO$_2$ using \textit{Spitzer} IRAC and observations of the forbidden oxygen lines in three comets: C/2012 K1 (PanSTARRS), C/2012 K5 (LINEAR), and 290P/J\"{a}ger, as well as observations of OH in PanSTARRS and J\"{a}ger.  Our measured CO$_2$ abundances are within the spread of values previously observed, corroborating previous observations that the typical CO$_2$/H$_2$O ratio in comets is in the range 15-30\% and consistent with the theory that CO$_2$ forms via grain surface reactions involving CO.  We find evidence for a possible extended source for H$_2$O sublimation in PanSTARRS, which we interpret as an icy grain halo, similar to that observed for C/2009 P1 (Garradd).  We detected all three forbidden oxygen lines only for PanSTARRS; for the other two comets only the 6300~\AA~and 6364~\AA~lines were detected.  Therefore for LINEAR and J\"{a}ger we only obtained an upper limit on the oxygen line ratio.  We compared the CO$_2$ abundance inferred from the oxygen line ratios to the CO$_2$ abundance observed by \textit{Spitzer} to evaluate our understanding of the photochemistry responsible for the release of \ion{O}{1} into the coma.  The upper limits derived for LINEAR and J\"{a}ger are not particularly constraining, but we determined that the empirical release rates from~\cite{McKay2015} reproduced the CO$_2$ abundance in PanSTARRS more accurately than the release rates from~\cite{BhardwajRaghuram2012}.  The reason why the empirical release rates seem to reproduce the CO$_2$ abundance more accurately than those determined from photochemical models like~\cite{BhardwajRaghuram2012} is unclear.  More work is needed on all fronts, observational, laboratory, and theoretical, to fully understand \ion{O}{1} emission in comets and employ it as a reliable proxy for CO$_2$.\\

\ack
We thank two anonymous reviewers whose comments improved the quality of this manuscript.  This work was supported by the NASA Planetary Atmospheres Program through grant number NNX08A052G.  This work is partially based on observations made with the \textit{Spitzer} Space Telescope, which is operated by the Jet Propulsion Laboratory, California Institute of Technology under a contract with NASA.  We thank the APO and Keck observing staff for their invaluable help in conducting the observations.  We are thankful to Matthew Knight for productive discussions concerning the coma morphology seen at optical wavelengths for C/2012 K1 (PanSTARRS), as well as David Schleicher, Michael Combi, and Erika Gibb for sharing their unpublished production rates.  We thank John Barentine, Jurek Krzesinski, Chris Churchill, Pey Lian Lim, Paul Strycker, and Doug Hoffman for developing and optimizing the ARCES IRAF reduction script used to reduce the ARCES data.  We would also like to acknowledge the JPL Horizons System, which was used to generate ephemerides for nonsidereal tracking of the comets during the ARCES observations, and the SIMBAD database, which was used for selection of reference stars.  The authors wish to recognize and acknowledge the very significant cultural role and reverence that the summit of Maunakea has always had within the indigenous Hawaiian community.  We are most fortunate to have the opportunity to conduct observations from this mountain.

\label{lastpage}


\bibliography{../references.bib}

\bibliographystyle{plainnat}

\clearpage

\begin{table}[h!]
\begin{center}
\caption{\textbf{Observation Log-\textit{Spitzer}}
\label{Spitzer}
}
\begin{tabular}{lccccc}
\hline
Comet & Date (UT) & R (AU) & $\Delta^a$ (AU) & Exp. Times (s) & Effective On-Source Exp. Time (s)\\
\hline
LINEAR & 1/31/2013 & 1.51 & 1.03 & 1.2 and 30 & 236\\
LINEAR & 2/15/2013 & 1.66 & 1.04 & 0.6, 12, and 100 & 936\\
J\"{a}ger & 2/3/2014 & 2.18 & 1.81 & 0.6, 12, and 100 & 562\\
PanSTARRS & 5/25/2014 & 1.83 & 1.24 & 0.6 and 6 & 26.4\\
\hline
\end{tabular}
\end{center}
$a$ Distance from \textit{Spitzer}\\
\end{table}

\clearpage

\begin{table}
\caption{\textbf{Observation Log-APO/Keck}
\label{observations}
\label{lasttable}
}
\begin{tabular}{lllllllllll}
\hline
Comet & Date (UT) & $r$ (AU) & $\Delta$ (AU) & $\dot{\Delta}$ (km s$^{-1}$) & G2V & Fast Rot. & Flux Cal\\
\hline
LINEAR & 2/7/2013 & 1.57 & 1.11 & 47.0 & HD 25370 & HD 27660 & HR 1544\\
LINEAR & 2/15/2013 & 1.66 & 1.35 & 46.2 & HD 25370 & HD 27660 & HR 1544\\
J\"{a}ger & 11/6/2013 & 2.45 & 1.78 & -23.7 & HD 259216 & HR 2532 & Hilt 600\\
J\"{a}ger & 11/15/2013 & 2.42 & 1.66 & -21.7 & HD 259516 & HR 2532 & Hilt 600\\
J\"{a}ger$^a$ & 1/14/2014 & 2.22 & 1.28 & 1.8 & Hyades 64 & HR 2207 & Hilt 600\\
PanSTARRS & 6/4/2014 & 1.71 & 1.69 & 19.5 & 35 Leo & 33 LMi & HD 93521\\
\hline
\end{tabular}
$^a$ Obtained with Keck HIRES
\end{table}

\clearpage

\begin{table}
\caption{\textbf{Observed Fluxes}
\label{fluxes}
\label{lasttable}
}
\begin{center}
\begin{tabular}{lllll}
\hline
Comet & Date (UT) & Species & Flux$^a$ & 3.6 $\mu$m Scale Factor$^b$\\
\hline
LINEAR & 1/31/2013 & CO$_2$ & 1.31 $\pm$ 0.36 & 3.25 $\pm$ 0.12\\
LINEAR & 2/15/2013 & CO$_2$ & 0.89 $\pm$ 0.18 & 2.59 $\pm$ 0.71\\
J\"{a}ger & 2/3/2014 & CO$_2$ & 1.55 $\pm$ 0.15 & 1.63 $\pm$ 0.24\\
PanSTARRS & 5/25/2014 & CO$_2$ & 38.8 $\pm$ 0.1 & 1.69 $\pm$ 0.01\\
LINEAR & 2/7/2013 & [\ion{O}{1}] & 0.39 $\pm$ 0.04 & -\\
LINEAR & 2/15/2013 & [\ion{O}{1}] & 0.30 $\pm$ 0.04 & -\\
J\"{a}ger & 11/6/2013 & [\ion{O}{1}] & 0.22 $\pm$ 0.03 & -\\
J\"{a}ger & 11/15/2013 & [\ion{O}{1}] & 0.33 $\pm$ 0.05 & -\\
PanSTARRS & 6/4/2014 & [\ion{O}{1}] & 23.8 $\pm$ 2.5 & -\\
J\"{a}ger & 1/14/2014 & OH & 5.97 $\pm$ 0.6 & -\\
PanSTARRS & 5/7/2014 & OH & 63500 $\pm$ 60 & -\\
\hline
\end{tabular}
\end{center}
$^a$ Fluxes are in 10$^{-15}$ ergs s$^{-1}$ cm$^{-2}$.  For CO$_2$ \textit{Spitzer} observations, fluxes are for a 33-pixel aperture.  For \textit{Swift} OH observations, flux is given for a 50-pixel aperture.  For [\ion{O}{1}] and OH from ARCES and HIRES, fluxes are integrated over the entire slit.\\
$^b$ Only applicable to \textit{Spitzer} observations of CO$_2$.
\end{table}

\clearpage
\begin{table}
\begin{center}
\caption{\textbf{Production Rates, CO$_2$/H$_2$O Ratios, and Collisional Quenching Factors}
\label{QCO2}
\label{lasttable}
}
\begin{tabular}{llllll}
\hline
 & \multicolumn{2}{c}{$Q$ (10$^{26}$ mol s$^{-1}$)} &\\
Comet & R (AU) & CO$_2$ & H$_2$O & CO$_2$/H$_2$O ($\%$) & Coll. Quench. Factor\\
\hline
PanSTARRS & 1.83 & 54.6 $\pm$ 0.1 & 435 $\pm$ 44$^a$ & 12.6 $\pm$ 1.3 & 1.25\\
PanSTARRS$^b$ & 2.04 & - & 950 $\pm$ 80 & - & -\\
LINEAR & 1.51 & 1.12 $\pm$ 0.08$^c$ & 3.88 $\pm$ 0.4 & 28.9 $\pm$ 3.6$^c$ & 1.03\\
J\"{a}ger$^d$ & 2.42 & 4.23 $\pm$ 0.37 & 13.5 $\pm$ 1.4 & 31.3 $\pm$ 4.2 & -\\
J\"{a}ger$^e$ & 2.18 & - & 10.2 $\pm$ 1.5 & - & 1.00\\
\hline\\
\end{tabular}
\end{center}
$a$ H$_2$O Production from ARCES observations of [\ion{O}{1}] emission.\\
$b$ H$_2$O Production from \textit{Swift} observations of OH.\\
$c$ Due to uncertainties associated with the model-dependent dust subtraction, these values may be better interpreted as upper limits.\\
$d$ Q$_{H_2O}$ from January Keck HIRES observations of OH, no collisional quenching factor is given due to no \ion{O}{1} observation being obtained at this epoch.\\
$e$ Q$_{H_2O}$ from November ARCES observations of [\ion{O}{1}]6300 emission.  Q$_{CO_2}$ is not provided as the \textit{Spitzer} observation was in early February, therefore a comparison to the H$_2$O production rate from November is not necessarily meaningful.\\
\end{table}

\clearpage

\begin{table}
\caption{\textbf{Inferred vs. Measured CO$_2$/H$_2$O Ratio}
\label{comparerates}
\label{lasttable}
}
\begin{center}
\begin{tabular}{llllll}
Comet & \ion{O}{1} Ratio &\multicolumn{4}{c}{CO$_2$/H$_2$O (\%)}\\
\hline
 & & BR12 & McKay2015A & McKay2015B & \textit{Spitzer} \\
\hline
PanSTARRS & 0.054 $\pm$ 0.002 & 3.7 $\pm$ 0.3 & 9.5 $\pm$ 0.4 & 14.4 $\pm$ 0.6 & 12.6 $\pm$ 1.3\\
LINEAR & $<$ 0.169 & $<$ 27 & $<$ 42 & $<$ 64 & 28.9 $\pm$ 3.6\\
J\"{a}ger & $<$ 0.247 & $<$ 53 & $<$ 77 & $<$ 116 & 31.3 $\pm$ 4.2\\
\hline
\end{tabular}
\end{center}
\end{table}

\clearpage
\begin{table}
\begin{center}
\caption{\textbf{\ion{O}{1} Release Rates}
\label{Wrates}
\label{lasttable}
}
\begin{tabular}{lllll}
\hline
Parent & \ion{O}{1} State$^a$ & \multicolumn{3}{c}{Release Rates (10$^{-8}$ s$^{-1}$)}\\
\hline
 & & McKay2015A & McKay2015B & BR2012\\
\hline
H$_2$O & $^1$S & 0.64 & 0.64 & 2.6\\
H$_2$O & $^1$D & 84.4 & 84.4 & 84.4\\
CO$_2$ & $^1$S & 50.0 & 33.0 & 72.0\\
CO$_2$ & $^1$D & 75.0 & 49.5 & 120.0\\
CO & $^1$S & 4.0 & 4.0 & 4.0\\
CO & $^1$D & 5.1 & 5.1 & 5.1\\
\hline
\end{tabular}
\end{center}
$a$ These rates are for a given electron state, not the line.  Therefore if not all lines coming from that state are observed, the branching ratio needs to be accounted for.  For $^1$D, both the 6300~\AA~and 6364~\AA~lines are usually observed, so no correction is needed.  However, for $^1$S, typically only the 5577~\AA~line is observed (as is the case in this work), so the above rates need to be multiplied by a branching ratio of 0.9 to get the yield for $^1$S atoms that will decay through the 5577~\AA~line.
\end{table}

\clearpage

\begin{center}
 Figure Captions
\end{center}

\begin{figure}[p!]
\begin{center}
\includegraphics[width=0.8\linewidth]{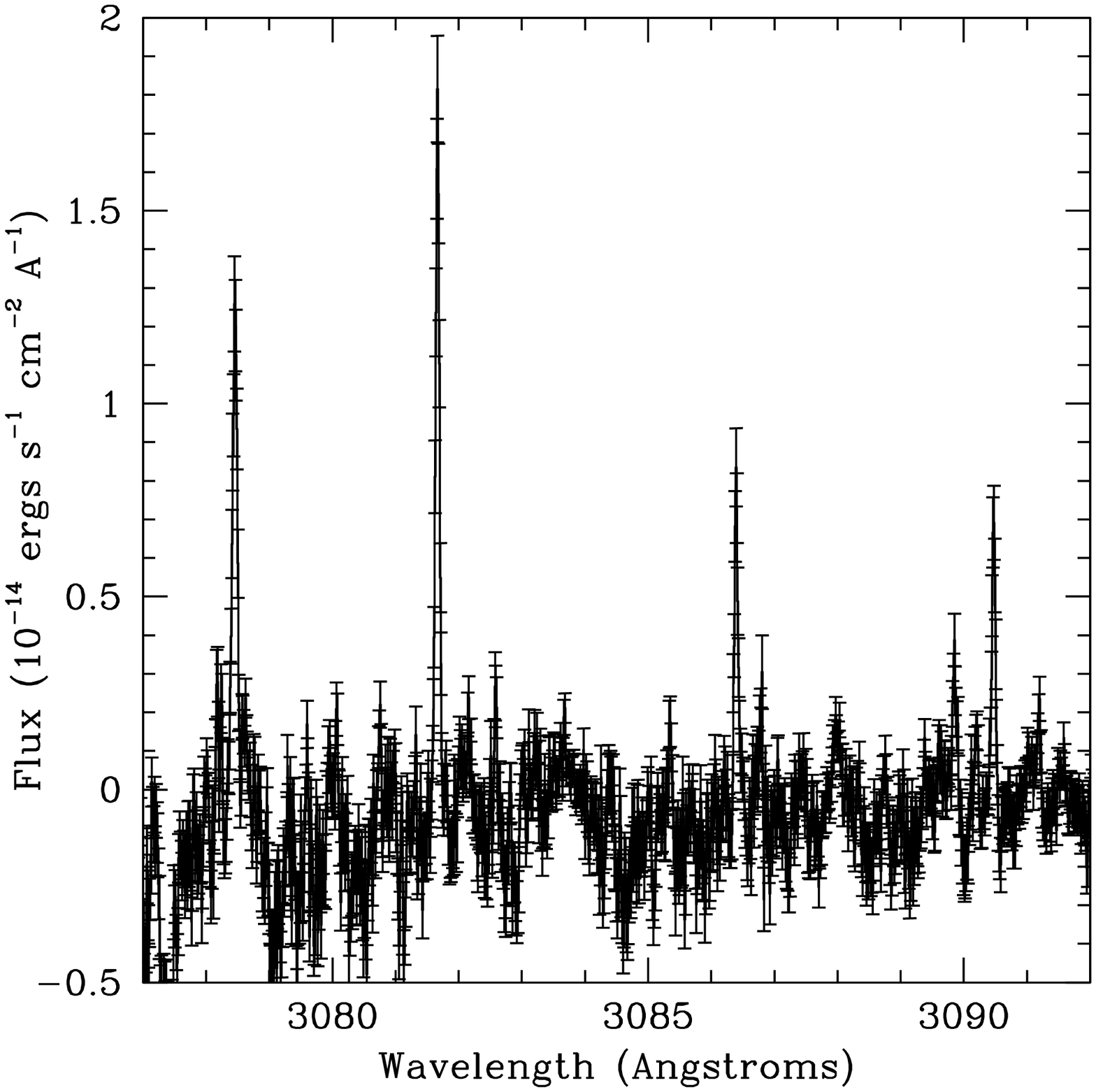}
\caption{
\label{OH}
}
\end{center}
\end{figure}

Fig~\ref{OH}: Spectrum of J\"{a}ger taken with Keck HIRES showing part of the $\Delta$v=0 OH band.  The background is due to the effect of scattered moonlight.

\begin{figure}[p!]
\begin{center}
\includegraphics[width=0.9\linewidth]{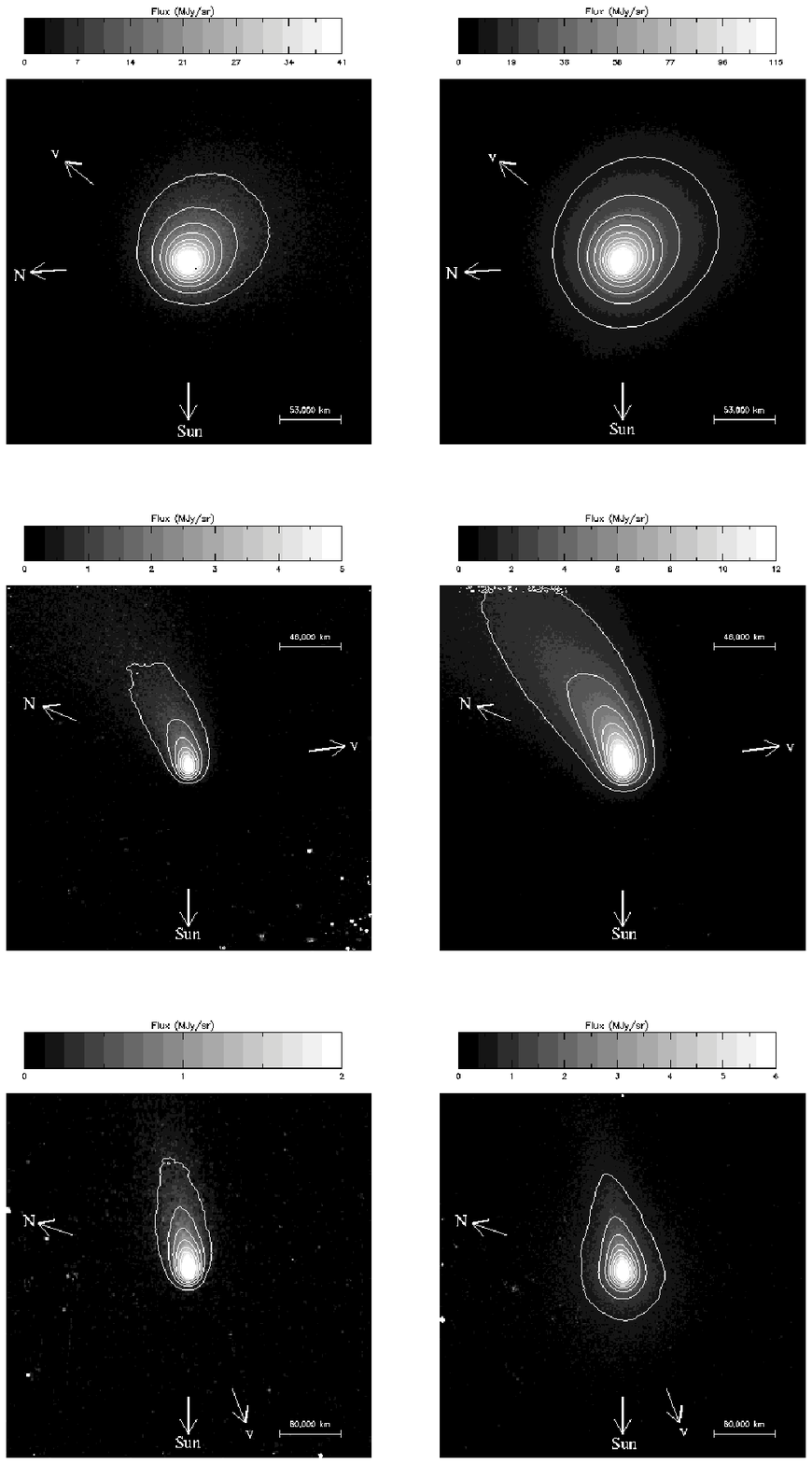}
\caption{
\label{Spitzer_images}
}
\end{center}
\end{figure}

Fig~\ref{Spitzer_images}: \textit{Spitzer} IRAC images at 3.6 (left column) and 4.5 (right column) $\mu$m of PanSTARRS (top row), LINEAR (middle row), and J\"{a}ger (bottom row).  The solar and velocity directions are indicated by the arrows labeled ``Sun'' and ``v'', respectively, while celestial north is depicted by the arrow labeled ``N''.  It is apparent in the 4.5 $\mu$m image of PanSTARRS that there is diffuse emission not present in the 3.6 $\mu$m image, which is likely due to CO$_2$.  There appears to be some diffuse emission in the 4.5 $\mu$m image of J\"{a}ger as well, though it is not as obvious as for PanSTARRS.  The gas emission is not obvious in the 4.5 $\mu$m image of LINEAR.

\begin{figure}[h!]
\begin{center}
\includegraphics[width=0.45\linewidth]{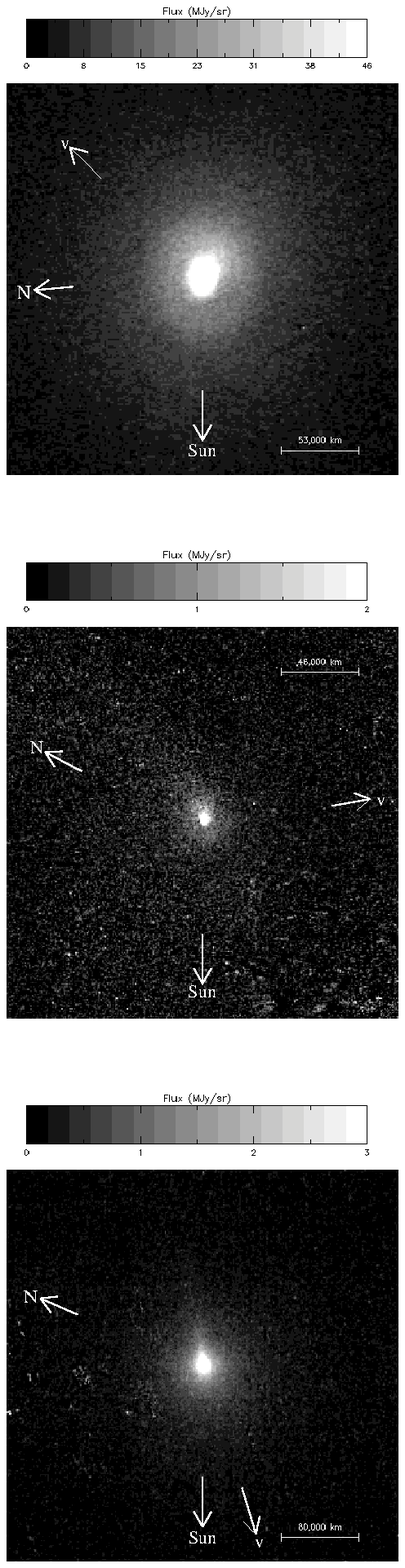}

\caption{
\label{Spitzer_gas}
\label{lastfig}
}
\end{center}
\end{figure}

Fig~\ref{Spitzer_gas}: Dust-subtracted images of PanSTARRS (top), LINEAR (middle), and J\"{a}ger (bottom).  The solar and velocity directions are indicated by the arrows labeled ``Sun'' and ``v'', respectively, while celestial north is depicted by the arrow labeled ``N''.

\begin{figure}[p!]
\begin{center}
\includegraphics[width=0.8\linewidth]{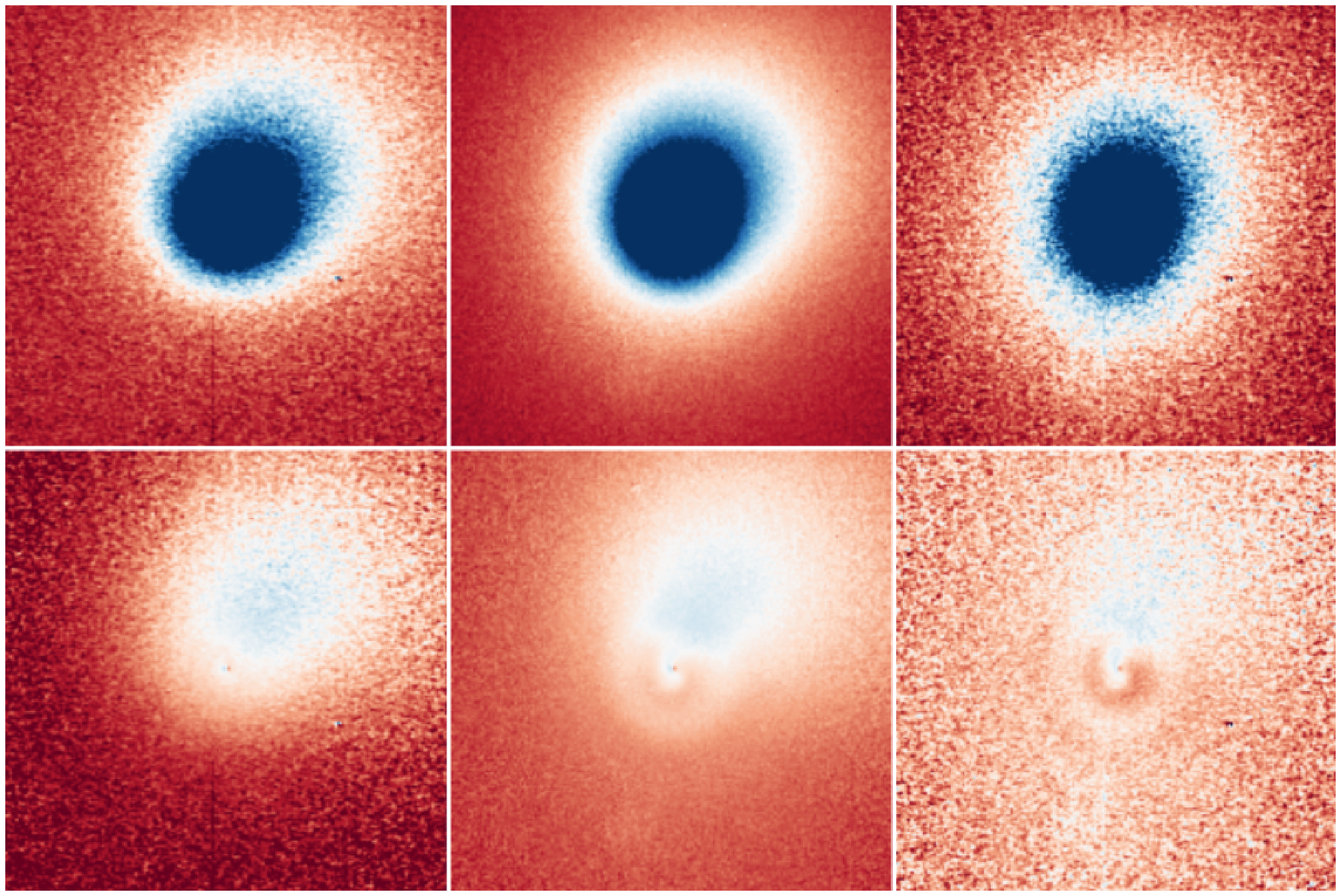}
\caption{
\label{PanSTARRS_morph}
}
\end{center}
\end{figure}

Fig~\ref{PanSTARRS_morph}: \textit{Spitzer} images of PanSTARRS before (top row) and after (bottom row) division by a 1/$\rho$ profile.  Left to right is 3.6 $\mu$m, 4.5 $\mu$m, and the dust-subtracted image.  The tail is obvious in the 3.6 $\mu$m and 4.5 $\mu$m images, and a faint residual is still evident in the dust-subtracted image.  The dust-subtracted and 4.5 $\mu$m images show a spiral structure that is not evident at 3.6 $\mu$m, which is likely the manifestation of a CO$_2$ jet.  Each subpanel has dimensions of 220,000 km on a side.

\begin{figure}[p!]
\begin{center}
\includegraphics[width=0.8\linewidth]{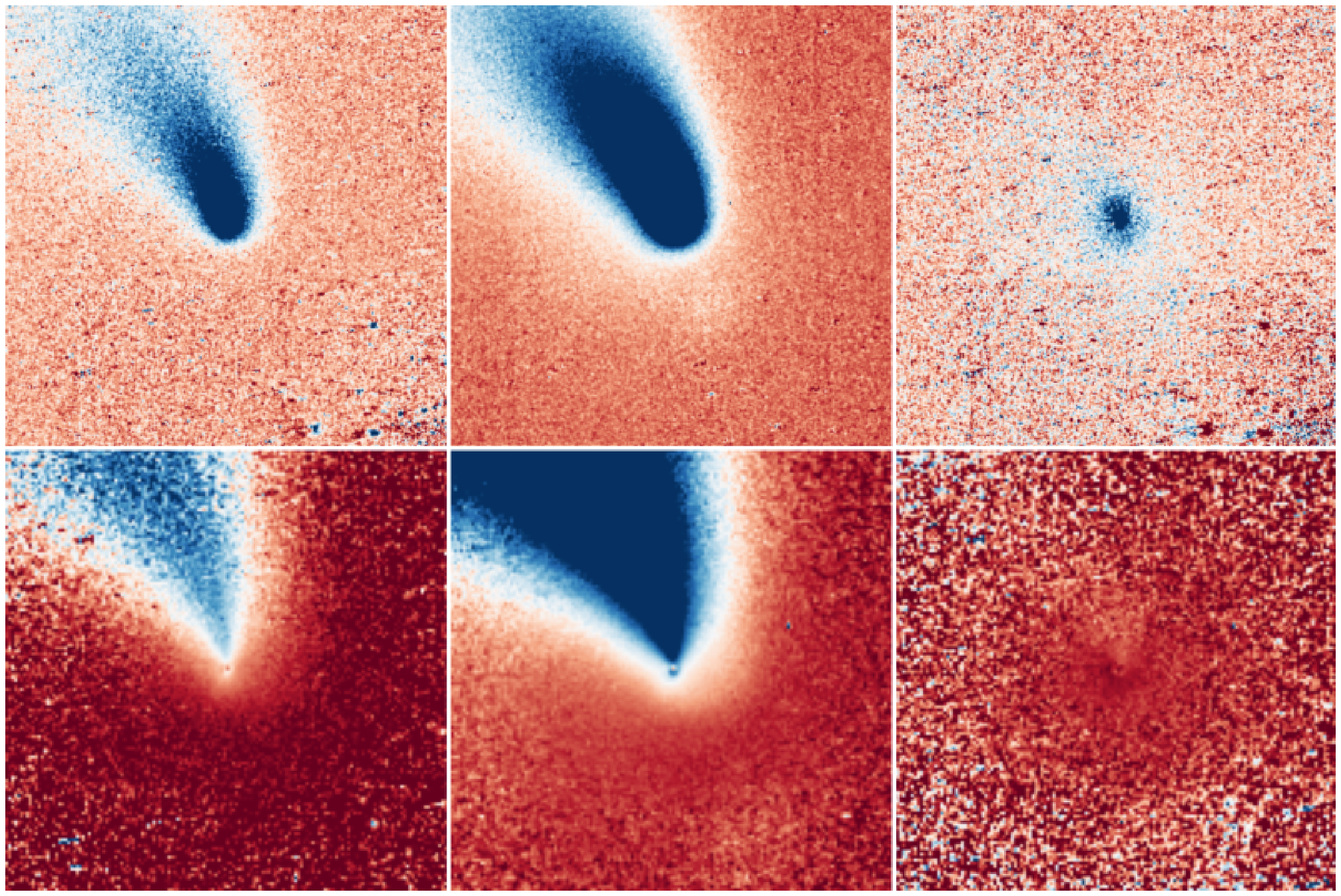}
\caption{
\label{LINEAR_morph}
}
\end{center}
\end{figure}

Fig~\ref{LINEAR_morph}: \textit{Spitzer} images of LINEAR before (top row) and after (bottom row) division by a 1/$\rho$ profile.  Left to right is 3.6 $\mu$m, 4.5 $\mu$m, and the dust-subtracted image.  The tail is obvious in the 3.6 $\mu$m and 4.5 $\mu$m images, and a faint residual is still evident in the dust-subtracted image.  In all images the coma morphology is symmetric, showing no obvious structures.  Each subpanel has dimensions of 140,000 km on a side.

\begin{figure}[p!]
\begin{center}
\includegraphics[width=0.8\linewidth]{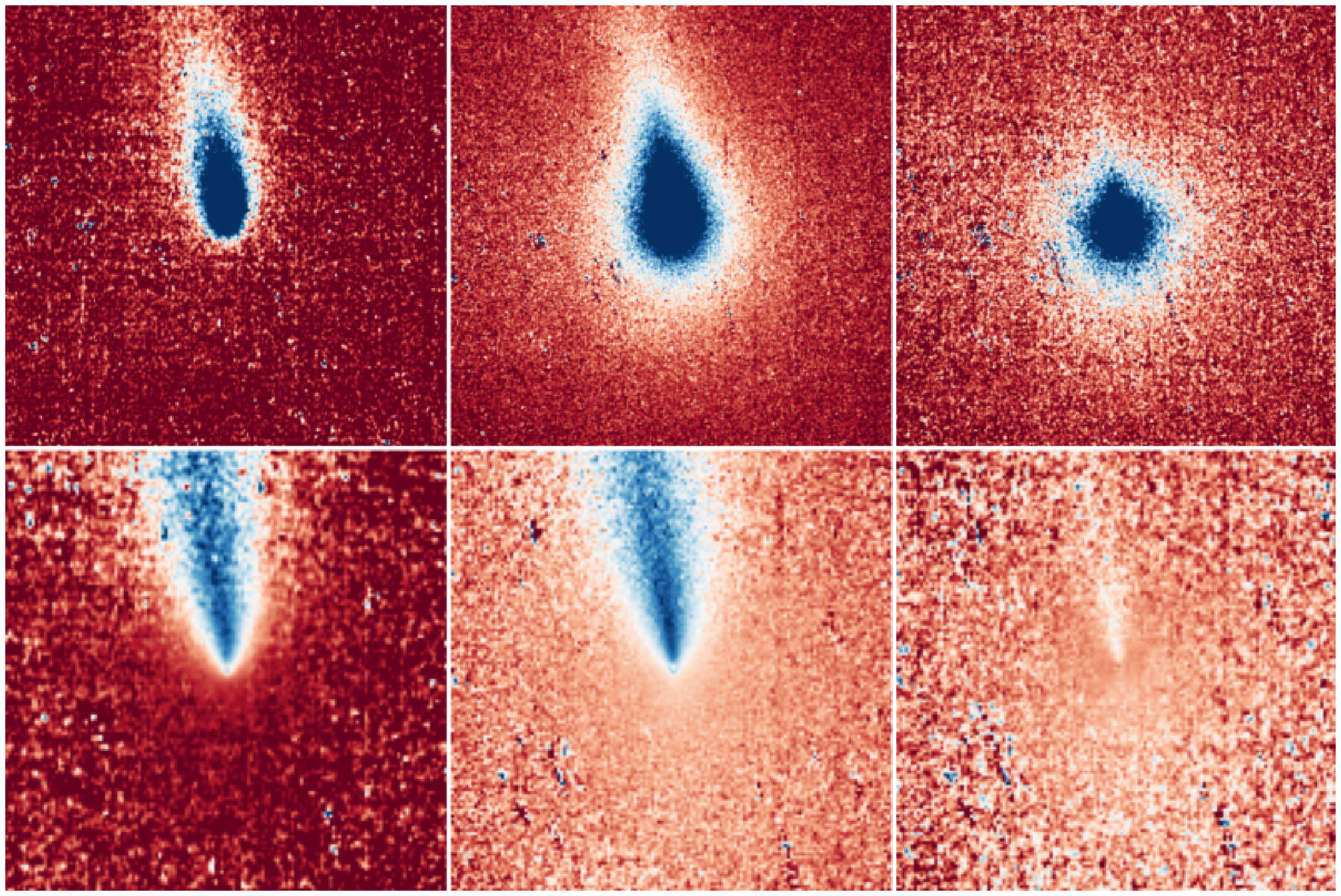}
\caption{
\label{Jager_morph}
}
\end{center}
\end{figure}

Fig~\ref{Jager_morph}: \textit{Spitzer} images of J\"{a}ger before (top row) and after (bottom row) division by a 1/$\rho$ profile.  Left to right is 3.6 $\mu$m, 4.5 $\mu$m, and the dust-subtracted image.  The tail is obvious in the 3.6 $\mu$m and 4.5 $\mu$m images, and a faint residual is still evident in the dust-subtracted image.  There is some possible extension of flux toward the bottom of the frame in the dust-subtracted image, but otherwise no coma features are present.  Each subpanel has dimensions of 240,000 km on a side.

\begin{figure}[p!]
\begin{center}
\includegraphics[width=0.8\linewidth]{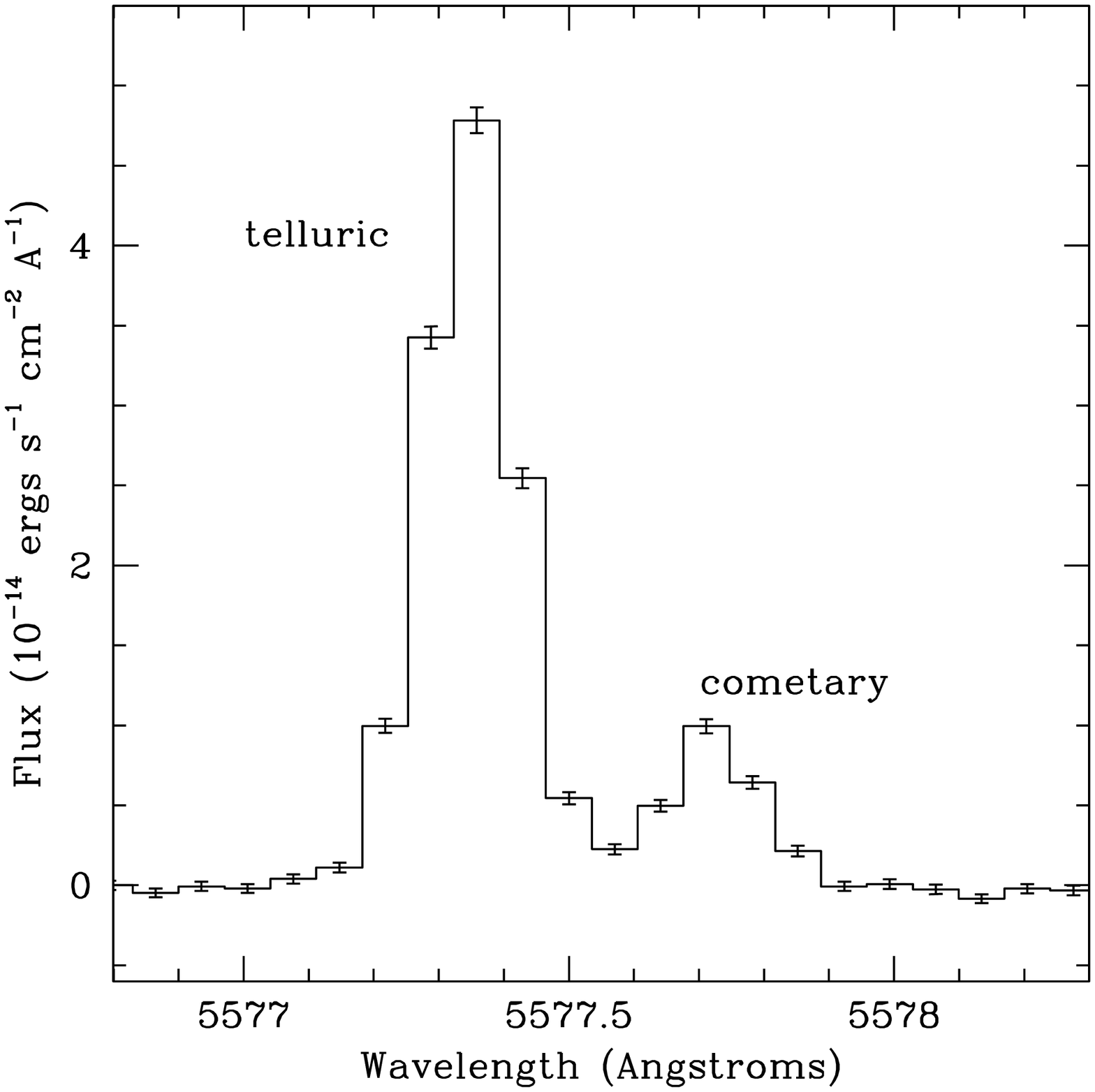}
\caption{
\label{OIline}
\label{lastfig}
}
\end{center}
\end{figure}

Fig~\ref{OIline}: Spectrum of PanSTARRS depicting the [\ion{O}{1}]5577~\AA~line.  The cometary line is redshifted compared to the telluric line and is significantly weaker in intensity.

\clearpage

\clearpage

\end{document}